# *Compact Variable-Gap Undulator with Hydraulic-Assist Driver: undulator harmonic energy scan test*

Alexander Temnykh[1*] and Ivan Temnykh [2]

[1]CHESS, Cornell University, Ithaca, NY 14850, USA;

[2] Pine Hollow Auto Diagnostics, Pennsylvania Furnace, PA 16865, USA

* Corresponding author, E-mail: abt6@cornell.edu

## Abstract

In the compact variable-gap undulator with Hydraulic-Assist Driver, miniature hydraulic cylinders compensate for ~90% of magnetic forces and mechanical actuators are used to accurately control the undulator gap. This method of hydraulic assist enables design simplicity and compactness of the undulator. Proof-of-concept experiments and test results of the undulator prototype in **step-by-step motion** mode indicated excellent accuracy and stability properties of the driver.

In this note we present test results of hydraulic-assist driver in **continuous motion** mode. Results indicate that hydraulic-assist driver can provide smooth continuous undulator gap change, keeping the undulator harmonic energy variation within the required 1eV/sec, and speed and RMS normalized energy errors at $1.8 \times 10^{-4}$ level (or less). That satisfies any application planned at CHESS.

## 1. Introduction

The present test was motivated by "Super" CCU (S-CCU) development project. The project itself was started at CHESS under "CHEX R&D Technology" program in 2019 with the goal to further improve CCU technology. It assumed development of a compact undulator with *variable gap controlled by hydraulic-assist driver and with hybrid magnetic structure with boosted magnetic field*. Both the hydraulic-assist gap driver and hybrid magnetic structure with boosted field, are novel. While hydraulic-assist driver was described in [1] and [2], magnetic structure of S-CCU will be described in future notes.

Concept of hydraulic-assist driver and the first proof-of-principle experiments were presented in [1]. In [2] the driver was employed in the compact variable gap undulator for precise gap control. In both references, the driver revealed excellent characteristics satisfying the most demanding applications in *step-by-step gap variation* mode.

The presented note describes additional test results of the driver operating in *continuous gap variation mode*. Continuous gap adjustment is required for undulator harmonic energies scanning. Evaluation was made using one of the existing CCUs converted to a variable gap undulator. Bench instrumentation and setup were similar to described in [2]. Results are presented below.

## 2. Instrumentation and Method

To control the driver (and undulator gap) we used interface based of LabView software. Description of the interface operation can be found in [2]. In order to provide continuous undulator gap adjustment, we added the option to allow smooth variation of the undulator gap "target" [2] with time.

Three signals were recorded by data acquisition system during gap variation:1) undulator gap "target"; 2) undulator gap "actual" acquired by measuring undulator girder positions with 0.5e-3 mm resolution optical encoders; 3) magnetic field measured with Hall sensor placed in the middle of undulator at maximum magnetic field location.

Fig. 2 displays example of the recorded data. The left plot shows undulator gap "target" and "actual" vs time. Two vertical lines indicate the time interval (ROI) used for the data analysis. Plot in the middle shows Hall sensor signal converted to magnetic field, and plot on the right displays first undulator harmonic energy (E1) as function of time in ROI. Harmonic energy was calculated using magnetic field data (middle plot).

In this particular example, in the beginning the gap "target" (blue line) was kept constant at 13.073 mm for 1 minute, then it was ramped to 14.120 mm in 11.66 minutes with constant 1.5e-3 mm/sec speed. After the ramp, it was changed back to 13.073 mm in one step and was kept there for another 3 minutes. The "actual" gap (red line) followed the "target" with few micron precision during constant speed ramp in ROI. When gap "target" was changed quickly at the 12 minute marker, the difference between gap "target" and "actual" became larger.

### 2.1. Undulator Gap variation and Undulator Energy scan characteristics.

According to CHESS beam scientists, the typical speed of harmonic energy change during undulator energy scan is ~1eV/sec or less. This speed defines undulator the gap change rate.

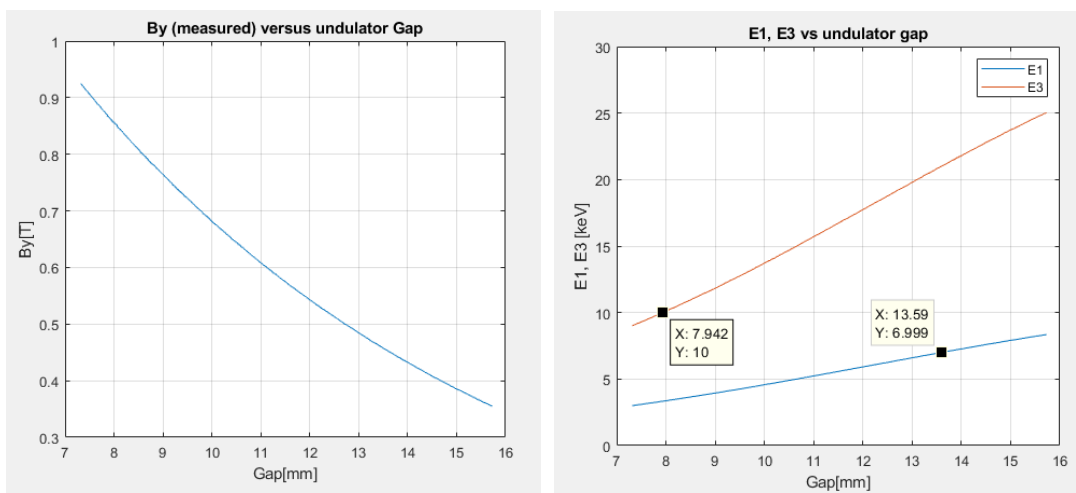

*Figure 1. On the left - peak magnetic field in the middle of undulator as a function of undulator gap. On the right - first and third harmonic energies as function of gap.*

In our case, the required gap variation rate was found in the following way: First, we measured the dependence of a peak magnetic field on gap (left plot on Fig.1), varying gap from the minimum 7.2mm to 15.7mm. Then the magnetic field data was translated into the first ($E_1$) and the third ($E_3$) undulator harmonic energies using formulas [3]:

$$E_n = 0.95 \frac{nE_b^2}{l_p(1 + K^2/2)}; K = 0.93 \times l_p \times B_{peak}$$

Where: $E_n$- undulator harmonic energy in $keV$; $n$ – harmonic order; $l_p$- undulator period in $cm$; $E_b$ – beam energy in $GeV$; $B_{peak}$ – undulator magnetic field peak amplitude. Dependence of $E_1$ and $E_3$ on undulator gap is shown on the right plot in Fig.1. Data indicates that when undulator gap is varying from minimum 7.2 mm to maximum 15.7 mm, $E_1$ changes from 3kev to 8.3 keV and $E_3$ from 9 keV to 25 keV.

For evaluating the performance of the undulator energy scan, two photon energy ranges, ~ 7 keV and ~10 keV, were selected. The first (~7keV) range is in the E1 variation span; the second (~10keV) is in the E3 variation span. Data on the right plot in Fig.1 indicated that the change rate of E1 and E3 with gap variation at E1=7 keV and E3 = 10 keV is ~0.667 keV/mm and ~1.667 keV/mm respectively. To provide required 1eV/sec rate of harmonic energy change, undulator gap variation speed should be 1.50e-3mm/sec and 0.601e-3 mm/sec.

## 3. Test results.

Fig.2 and Fig.3 illustrate E1 scanning around 7 keV and E3 scanning around 10 keV.

*E1 scanning around 7 keV.* For a 28.4mm undulator period and 6GeV beam energy, E1 equal to 7 keV at 4.520 kG peak field required an undulator gap of 13.59 mm(left plot in Fig.1). To provide 1eV/sec energy change rate, the gap variation speed should be 1.50e-3 mm/sec and to cover ~10% of the energy span, the gap should be changed by 1.047 mm.

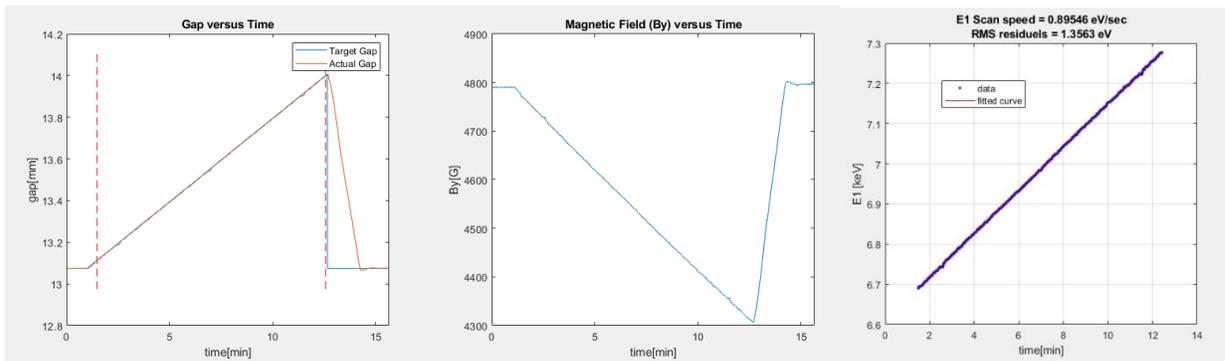

*Figure 2, E1 scan from 6.65 to 7.3 keV. Harmonic energy and undulator gap variation rates are 0.91 eV/sec and 1.50e-3 mm/sec. On the left – gap versus time. In the middle – peak magnetic field in the undulator center. On the right – the time dependence of first undulator harmonic energy calculated using peak magnetic field and the fit.*

Undulator gap "target" dependence on time is shown by a blue line on the left plot in Fig.2. It was changed by 1.047 mm around 13.56 mm with 1.50e-3 mm/sec speed. The "actual" gap(red line) followed the "target" with high accuracy during smooth ramping in "ROI".  Plot in the middle shows magnetic field varying around required 4.520 kG and plot on the right displays the E1 dependence on time in "ROI" calculated using magnetic field data. 7 keV energy is in the middle of E1 variation span.

Second order polynomic fit of the E1 dependence on time yields ~0.90 eV/sec changing rate, which is close to required 1 eV/sec, and ~1.36 eV RMS of residual errors. These errors can be attributed to errors in the undulator girder positioning, i.e. driver inaccuracy and to the Hall sensor noise. Magnetic field measurement at constant gap revealed 0.82 Gauss RMS noise. That translates to 1.02 eV RMS in E1 variation. Assuming that magnetic field noise and errors due to driver inaccuracy are uncorrelated, we can estimate effect of latter on energy variation error as 0.90 eV RMS or 1.3e-4 of normalized energy. Knowing how E1 depends on gap, we assess the gap (driver) inaccuracy as 0.90 eV/(667 eV/mm) ~ 1.3e-3 mm.

***E3 scanning around 10 keV*** required undulator peak field varying around 8.618 kG and gap varying around 7.942 mm. For 1 eV/sec E3 scan rate, the gap changing rate should be 0.601e-3 mm/sec and for 10% energy span required 0.60 mm of total gap change.

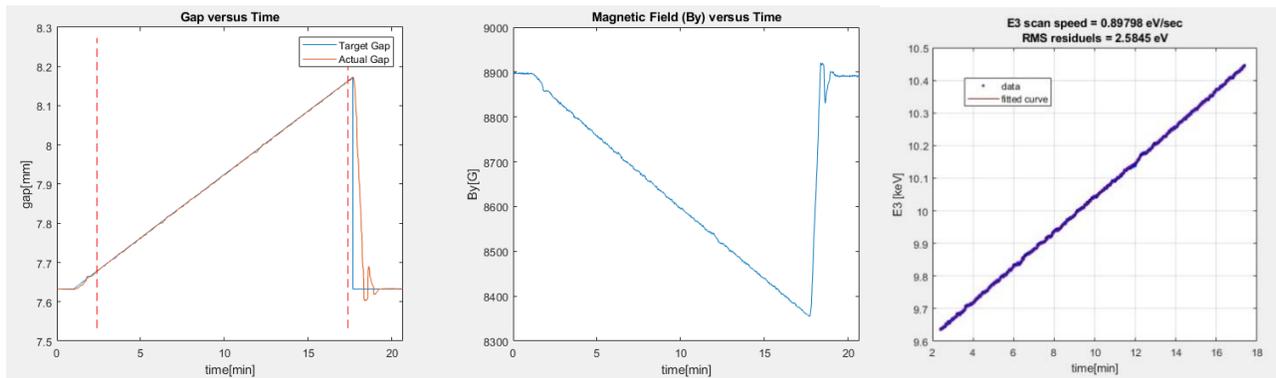

*Figure 3. Third harmonic (E3) scan from 9.6 to 10.44 keV. Undulator gap variation speed 0.60e-3 mm/sec provided harmonic energy variation rate 0.90 eV/sec. On the left – undulator gap as a function of time. Dashed lines indicate region of interest (ROI). In the middle – peak magnetic field measured in the undulator center. On the right – the time dependence of first undulator harmonic energy calculated using peak magnetic field.*

Fig. 3 illustrates the scan data. Left plot shows gap "target", "actual" gap versus time and ROI. Plot in the middle is a peak magnetic field and plot on the right is a third harmonics energy (E3) in ROI.

The fit of E3 dependence on time yields ~0.90 eV/sec changing rate and ~2.58 eV RMS of residual errors. 1.78 eV RMS of these errors was attributed to Hall sensor nose. The rest 1.86 eV RMS (~1.86e-4 of RMS normalized errors) is due to gap (driver) inaccuracy. The gap (driver) positioning errors were estimated as 1.86 eV/(1667 eV/mm) ~ 1.1e-3 mm. It consists with results obtained from E1 scanning around 7 keV.

## 4. Conclusion.

In continuous gap variation (undulator energy scan) mode, the Hydraulic-Assist Driver demonstrated a gap control accuracy of ~1.3e-3 mm (or better). This results in ~1.86e-4 (or smaller) of RMS normalized harmonic energy errors which is ***~100x*** smaller than harmonics FWHA.

As a conclusion we state that compact variable gap undulator with Hydraulic-Assist Driver will satisfy the most demanding applications required at stationary as well as continuously-varying undulator energies.

## Acknowledgment

The Work was supported by the NSF award DMR-1829070 and by the efforts from Pine Hollow Auto Diagnostics.

Authors would like thank CHESS ID2A and ID4B beam line scientists  Christopher Pollock and Jacob Ruff for motivation and help in selection of scanning parameters.